**Who discovered Universe expansion?**

Controversy persists over who first found that the Universe is expanding. Last year, Mario Livio quashed suggestions that Georges Lemaître's 1927 theoretical prediction of expansion was deliberately suppressed (*Nature* **479,** 171–173; 2011). Since then, another contender has emerged.

The joint NASA and Infrared Processing and Analysis Center Extragalactic Database of Galaxy Distances, in Pasadena, California, which I co-lead, has tabulated and made public the historical distance estimates published by Edwin Hubble and his contemporaries to prove expansion (see I. Steer *J. R. Astron. Soc. Can.* **105,** 18–20; 2011). These reveal that measurements by a Swedish astronomer, Knut Lundmark, were much more advanced than formerly appreciated.

Lundmark was the first person to find observational evidence for expansion, in 1924 — three years before Lemaître and five years before Hubble. Lundmark's extragalactic distance estimates were far more accurate than Hubble's, consistent with an expansion rate (Hubble constant) that was within 1% of the best measurements today.

However, Lundmark's research was not adopted because it relied on one unproven method (galaxy diameters), cross-checked with one unproven distance to the Andromeda galaxy, which was derived from a type Ia supernova observed in 1885 and mistaken for a normal nova (W. Huggins and W. F. Denning *Nature* **32,** 465–466; 1885).

Hubble's research in 1929 yielded a value for the Hubble constant that was inaccurate by almost an order of magnitude. It was adopted because it was derived from multiple methods — including one still in use (brightest stars) — and was cross-checked with multiple galaxies with distances based on proven Cepheid star variables.

Lundmark established observational evidence that the Universe is expanding. Lemaître established theoretical evidence. Hubble established observational proof.

**Ian Steer**
*NASA/IPAC Extragalactic Database of Galaxy Distances, Pasadena, California, USA.*
iansteer1@gmail.com

Further information will be presented in: Hubble's Law: Who Discovered What and When, at the American Astronomical Society meeting (AAS 221), in Long Beach, California, January 6-10, 2013, during the AAS Historical Astronomy Division (HAD) VI History of Astronomy session, Tuesday, January 8, 2013, 10:00 AM to 11:30 AM.